\documentclass[%
 aip,article,
 amsmath,amssymb,
 preprint,
 reprint%
]{revtex4-1}
\usepackage{graphicx}
\usepackage{dcolumn}
\usepackage{bm}
\usepackage[utf8]{inputenc}
\usepackage[T1]{fontenc}
\usepackage{mathptmx}
\usepackage{etoolbox}

\begin{document}

\title{Sub-Unit Cell Spatial Resolution of Phase and Polarization Mapping in Scanning Electron Diffraction}


\author{Tim B. Eldred}
\affiliation{Department of Materials Science and Engineering, North Carolina State University, Raleigh, NC, 27607, USA}
\affiliation{Analytical Instrumentation Facility, North Carolina State University, Raleigh, NC, 27607, USA}
\author{Jacob G Smith}
\affiliation{Department of Materials Science and Engineering, North Carolina State University, Raleigh, NC, 27607, USA}
\author{Wenpei Gao}
\affiliation{Department of Materials Science and Engineering, North Carolina State University, Raleigh, NC, 27607, USA}
\affiliation{Analytical Instrumentation Facility, North Carolina State University, Raleigh, NC, 27607, USA}

\begin{abstract}
Diffraction analysis in four dimensional scanning transmission electron microscopy now enables the mapping of local structures including symmetry, strain, and polarization of materials. However, measuring the distribution of these configurations at the unit cell level remains a challenge because most analysis methods require the diffraction disks to be separated, limiting the electron probe sizes to be larger than a unit cell. Here we show improved spatial resolution in mapping the polarization displacement and phases of BaTiO3 sampled at a rate equivalent to the size of the projected unit cells using 4DSTEM. This improvement in spatial resolution is accomplished by masking out the overlapping regions in partially overlapped CBED patterns. By reducing the probe size to the order of single projected unit cells in size, the fluctuations of symmetry and structure inside the rhombohedral symmetry nanoregions inside of tetragonal domains can be observed. With the mapping performed at the unit cell resolution the nano-sized rhombohedral domains break down with local fluctuation, showing different character at different length scales.
\end{abstract}

\maketitle

\section{\label{sec:level1}Introduction}

The structural configurations in materials include symmetry, phase, composition, and distribution of domains. They are described at the atomic, nano, micro, and macro length scales, depending on the characterization technique employed, and related to the materials properties and functions. Identifying the evolution of structure at different length scales plays an important role in connecting the structure with their properties and optimizing processes in materials synthesis and engineering. In complex oxide, for example, to understand the polarization, symmetry, and strain usually requires the application of multiple characterization methods. \cite{martin_thin-film_2016, kim_determination_2013, yadav_observation_2016, kim_response_2017} In BaTiO3, the first perovskite oxide that was shown to be ferroelectric at room temperature,\cite{hans_insulating_1947}  a combination of diffraction analysis, imaging, and simulation has been done and only recently, evidences show that the tetragonal phases of BaTiO3 at room temperature may possibly emerge from rhombohedral phases that exist in domains at nano scale. \cite{tsuda_two-dimensional_2013, shao_nanoscale_2017} The appearance of such nanoscale regions of rhombohedral symmetry with coordinated 002 polarization can be explained by the order disorder phase transition.\cite{stern_character_2004,polinger_ferroelectric_2013}    Beyond identifying the symmetry and phases in small dimensions that are different from the bulk materials phases, mapping the distribution of such structural configurations is also critical.  Recently, exotic states including polarization vortex, skyrmion and meron, are discovered in the thin films of complex oxide multi-layers and supper-lattices, where these new states, probed and confirmed by atomic resolution scanning transmission electron microscopy \cite{yadav_observation_2016,nelson_spontaneous_2011}, are used as input for theoretical calculations that show how the energy terms across different length scales are balanced and stabilize the new ferroelectric. states\cite{hong_stability_2017}   Despite the high precision, major characterization methods,  like X-ray diffraction, measure averaged structural information within the broad beam size, local structure features critical to the materials electronic and magnetic properties is lost. To identify the local structures, measurement needs to be done using probes with sizes close to the dimension of the local features.

In transmission electron microscopes (TEM), the beam sizes can be adjusted from the atomic scale continuously to the micro scale, making it possible to probe characteristics specific to different regions, such as the works done using scanning electron nanodiffraction \cite{tsuda_two-dimensional_2013, zuo_nanometer-sized_2001, yuan_lattice_2019} and four dimensional scanning transmission electron microscopy (4D STEM). \cite{ophus_four-dimensional_2019}  In regular STEM mode, an electron beam is converged into a electron probe with its size determined by the convergence angle and lens aberration. To take an image, the probe is raster scanned across the sample, convergent beam electron diffraction patterns form in the diffraction plane, and electrons scattered to different angles are collected by detectors and counted to yield a number as the pixel intensity.  By segmenting an annular STEM  detector into quadrants and using it to capture the CBED pattern, differential phase contrast (DPC) measures the difference of intensity falling on the segmented detectors which reflects the intensity shifts. \cite{shao_nanoscale_2017} Using this technique, the internal electric fields are revealed in a variety of materials systems, including GaAs P-N junctions and GaN/AlGaN/GaN hetero-structures.\cite{beyer_quantitative_2021, muller-caspary_electrical_2019} Atomic resolution DPC was also demonstrated on BaTiO3, SrTiO3 and on single gold atoms. \cite{shibata_differential_2012, shibata_electric_2017} Recently, with the development of high speed electron detectors, 4D STEM data can be collected by synchronizing the scanning beam in STEM mode with one of the primary cameras which collects the entire CBED pattern as a 2D image at each position of the probe, as shown in Fig. \ref{fig:masking}. As the interaction of the forward moving electrons with the sample takes places in regions limited by the probe size, the polarization and symmetry can be measured by quantifying their influence in the CBED pattern. DPC imaging in 4D STEM also enabled the mapping of local electron density and fields. \cite{sanchez-santolino_probing_2018, fang_atomic_2019, gao_real-space_2019} By reconstructing the sample phase object using ptychography, it is possible to improve the image resolution beyond the diffraction limit. \cite{wang_electron_2017, jiang_electron_2018}

With the advancement of 4D-STEM comes the chance for bridging the gaps between the dimensions of the measurement probes by easily changing the probe sizes while scanning the same sample region. However, there are also challenges: in terms of determining the atomic structure, especially the polarization and atom displacement, atomic resolution high angle annular dark field (HAADF) and iDPC images are 2D images which does not necessarily make the determination of 3D structure straightforward; while diffraction analysis might reveal the 3D structural symmetry, the spatial resolution is larger than several unit cells.  For example, to map the local phase and symmetry of BaTiO3, most diffraction experiments are performed using convergence angles below the Bragg angle to make the diffraction disks separated, analysis of the internal intensity in each diffraction disks is then possible. The relatively low convergence angle results in electron probe sizes larger than the area of multiple unit cells for most materials.  It thus requires new techniques in 4D-STEM to consider each unit cell specifically and reveal the structures at the single unit-cell level. 

To this end, we developed a method to analyze partially overlapping coherent CBED patterns and present the mapping of the polarization in BaTiO3 here. As the overlap of diffraction disks affords a probe size on the scale of an individual unit cell, the structural information is extracted at the single unit-cell level, which can be binned to represent the structural evolution across length scales. Together with insights from diffraction simulations, we demonstrate that in BaTiO3 the population of rhombohedral phase with opposite atom displacement fluctuates at the unit cell level within the rhombohedral domains at the nano scales.

\section{Method}
\subsection{Simulation Details}
Simulations of CBED were performed using the Bloch wave methods in EMAPS.\cite{zuo_web-based_2004} A series of CBED was produced with sample thickness from 0.4 to 40 nm. This process was repeated for cubic, tetragonal, centered tetragonal, and rhombohedral phases. \cite{ehses_temperature_1981}
Multi-slice method was applied to simulate 4D-STEM data across a unit-cell of BaTiO3 using MulTEM. \cite{lobato_multem_2015} The 4D datasets were then averaged over the entire unit cell to generate position averaged CBED (PACBED) to remove the dependence on probe position from the pattern results and compare with experimental PACBED to measure the sample thickness.\cite{lebeau_position_2010}

\subsection{Experimental Details}
STEM samples were prepared by thinning a BaTiO3 single crystal with the [001] in the plane of the wafer. Samples were wedge polished before ion milling using a Gatan PIPS II using argon under LN2 temperatures in order to achieve a final thickness in the 5-20 nm regime with suitable flatness. To avoid the tetragonal to cubic transformation, the samples were brought back from LN2 temperatures slowly and allowed to equilibrate for at least 24 hours at 20C before imaging.
4DSTEM Data sets were acquired using an aberration corrected FEI-Titan STEM operated at 200 kV outfitted with a 128 x 128 pixel EMPAD detector as the diffraction camera.\cite{tate_high_2016} 4DSTEM was collected in microprobe mode sing a 4 mrad semi-convergence angle with a 0.24 nm step size.

\begin{figure}[htp]
\centering
\includegraphics[width=3.2in]{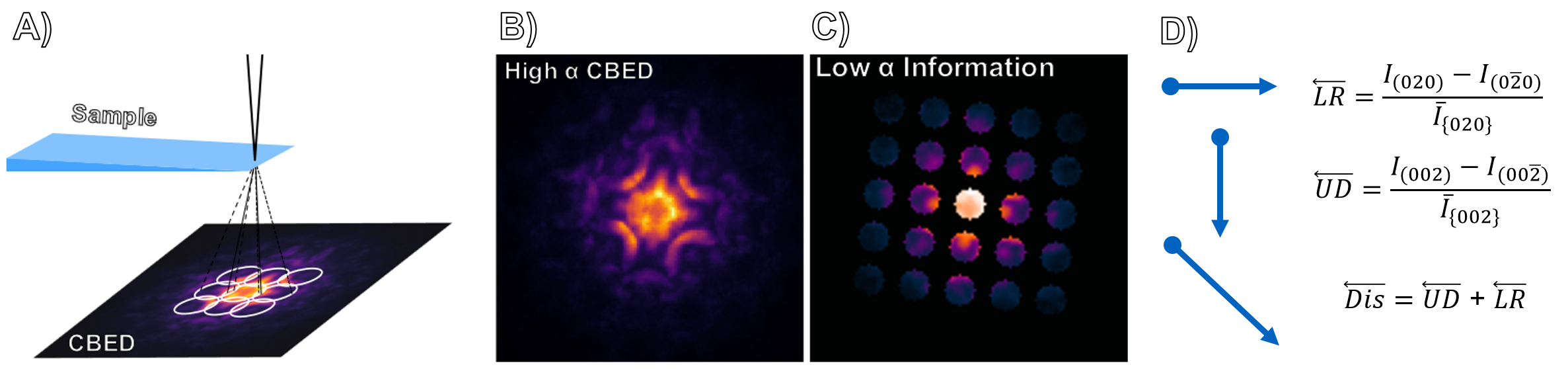}
\caption{A) CBED 4DSTEM Data acquisition scheme, with B) partially converging CBED patterns. C) Experimental patterns showing the retained low-semi convergence angle information. D) Vector equations for Friedel intensity analysis}
\label{fig:masking}
\end{figure}

\section{Results and Discussion}

\begin{figure}[htp]
\centering
\includegraphics[width=3.2in]{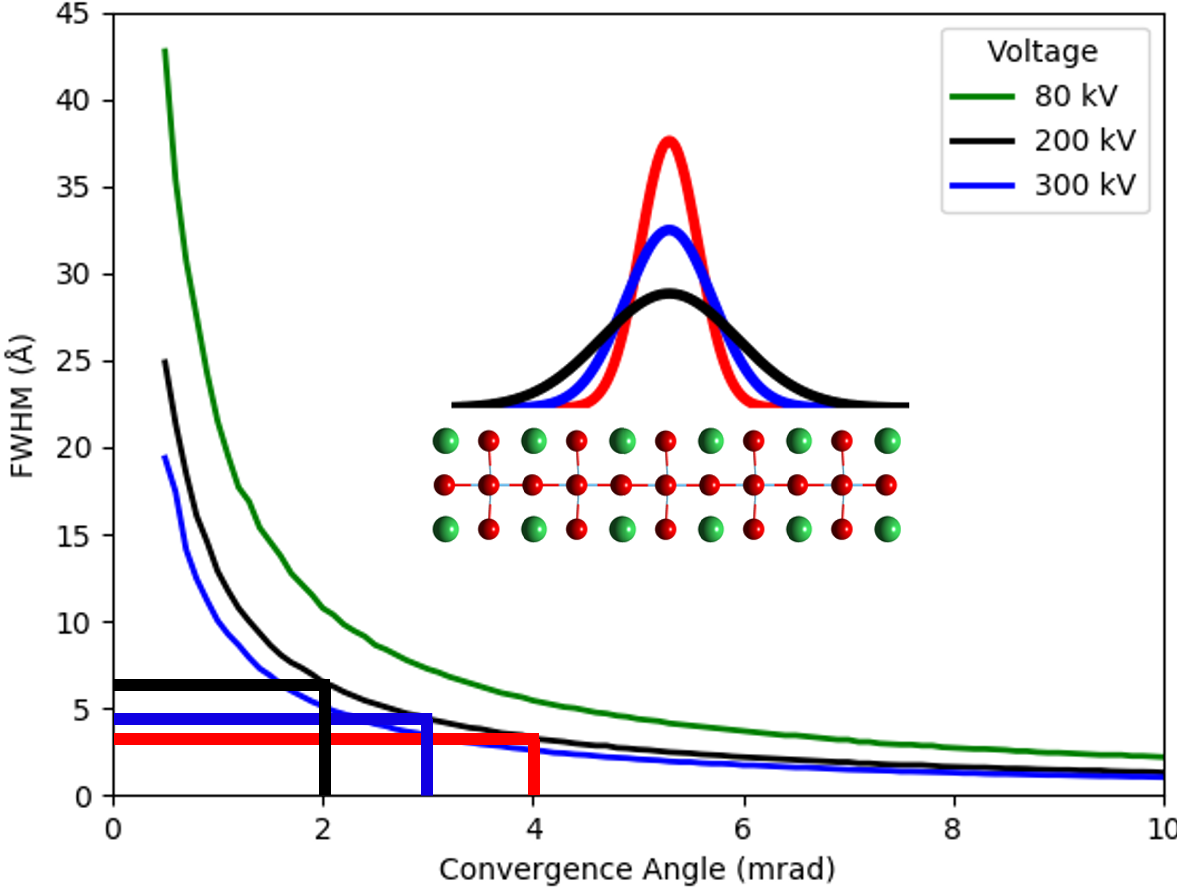}
\caption{MULTEM simulated probe sizes for low angle STEM.  (inset) Comparison of 2, 3, and 4 mrad probes to unit cell dimensions.}
\label{fig:probe}
\end{figure}

The sizes of the electron probe are calculated from simulated multislice datasets in figure \ref{fig:probe}. Using a convergence angle of 4 mrad, the full width at half maximum (FWHM) of the electron probe is 3.6 \AA, compared to the unit cell of BaTiO3 with 4.09x4.09 \AA in size. In contrast to a 7 \AA probe size using a convergence angle of 2 mrad, sampling more than 4 columns of unit cells at each probe position. 

We first compared the simulated diffraction data using 2 and 4 mrad as the semi-convergence angle and based on simple cubic, tetragonal, and rhombohedral phases. A mask for the diffraction pattern was applied to include only the interior 2 mrad around the center of each diffraction disc. By applying this mask at to each diffraction pattern as shown in Figure \ref{fig:probe}, each dataset (simulated and actual) was processed to contain only the non-overlapping regions of the patterns for analysis. When applying the masks to real data sets, positional-averaged diffraction patterns are used to determine the center position and radius of the non overlapping regions. Each pattern is then masked individually as shown in Figure \ref{fig:empad} and a new 4DSTEM data set generated from the masked data.

After normalizing the two datasets by the max intensity in the direct beam, the data sets were compared. The difference between the two patterns is on the order of less than 1\% difference, with no clear structure. This indicates that the differences are likely random and due to the uncertainty from frozen phonon configurations.  The symmetry of the patterns as well as the Friedel pair intensity ratios were maintained in the masked regions for all simulated configurations resulting in equivalent displacement vectors calculated from both the 2 and 4 mrad datasets.

The intensity in Friedel pairs of the CBED patterns was analyzed and compared with simulated data to quantify the atom displacement.\cite{tsuda_two-dimensional_2013} Using the formula in Figure \ref{fig:masking}.D the <002> (left to right, LR) and <020> (up to down, UD) components of the diffraction are used to calculate the polarization. From these two vectors, a displacement vector with magnitude, M, and angle, $\theta$, is computed and represents the overall polarization. We define $\theta$ as the deviation from the [002] direction to show the how the rhombohedral phase modifies the atom displacement in local region in relation with the overall tetragonal phase. The as-acquired Nanobeam Electron Diffraction (NBED) data were analyzed to show the distribution of polarization at the sub-unit cell scale, the 4D data sets were also averaged with 2x2 and 4x4 binning in position to map the polarization at larger scales. The polarization also influence the local electric field, which impacts the momentum of the electron probe as it propagates through the sample. A measurement of the center of mass (COM) of the intensity in the diffraction can reflect the momentum of the electron probe. The electric field was thus mapped using the COM in the masked CBED and compared with the polarization maps calculated using the Friedel diffraction pairs.

\begin{figure}[htp]
\centering
\includegraphics[width=3.2in]{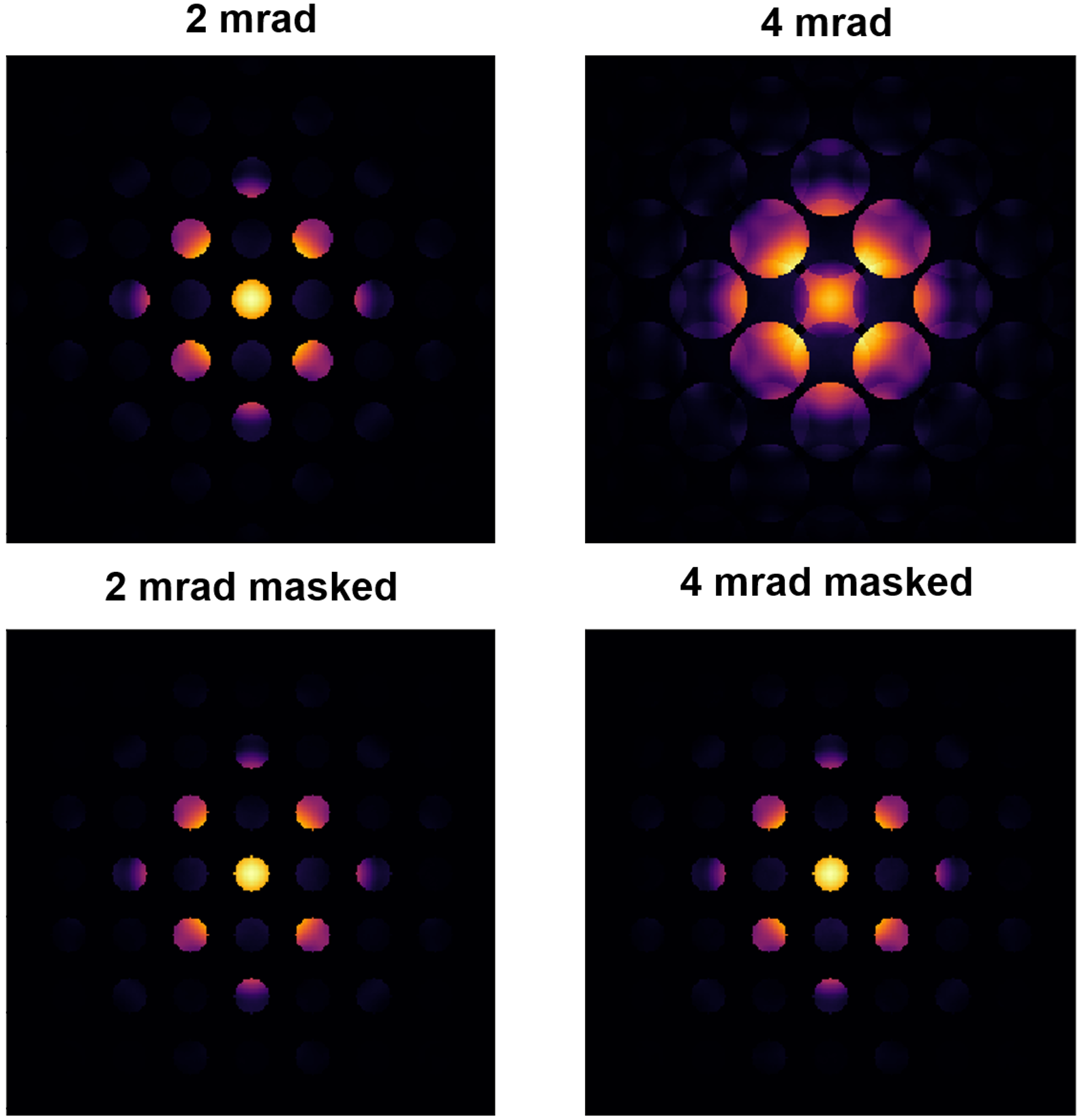}
\caption{Simulated data at 2 and 4 mrad convergence angles pre and post masking showing the equivalence of the structure of the masked data. Intensity Difference between the two patterns less than 1\%}
\label{fig:sim}
\end{figure}

For measurements of symmetry we considered two different approaches. A cross correlation symmetry analysis method was considered \cite{shao_nanoscale_2017}, however due to low dynamic diffraction due to the thin samples being considered, the R values for the tetragonal and rhombohedral phases were too similar to be differentiated. The method applied by Tsuda, et al., looks at intensity variations between pairs of diffraction spots across the mirror planes.\cite{tsuda_two-dimensional_2013} The break in the symmetry of the Friedel pair intensities is correlated to the break in symmetry of the overall unit cell. From the simulated patterns of the tetragonal and rhombohedral unit cells, we observed unequal intensity distribution across the second order Friedel pairs consistent with the overall unit cell symmetry. In the tetragonal phase, we observed unequal total intensity between the (002) discs, but the total intensity in the (020) discs is equivalent. In the cubic phase the total intensity in  all four {002} disks are the same, in agreement with 4-fold total symmetry. The rhombohedral structure as expected shows 2-fold symmetry in the diffraction pattern, with (002) and (020) discs possessing equal total intensity, as well as the (00-2) and (0-20) discs. In the simulated data, the magnitude of the displacement vector as shown in Figure \ref{fig:masking}  changes with thickness, however the polarization direction remains the same. Through comparison of the intensities of these key discs, the projected average displacement vector at the probe position can be determined for any given CBED pattern.

Experimentally, scanning electron diffraction data set was acquired from a BaTiO3 sample with a thickness of 19 nm. The representative diffraction patterns taken using the semi-convergence angle of 2 mrad and 4 mrad from locations of equal thickness are presented in Figure \ref{fig:empad}. By applying a mask of 2 mrad to the 4 mrad data, the diffraction disks and intensity distribution within are similar to the 2 mrad data. The deviation can be attributed to the slight difference in locations where each diffraction was taken. Using the experimental data, Friedel pair intensity ratios were analyzed and a phase and polarization map generated with the results presented in Figure \ref{fig:mapping}. This 4D dataset was obtained at a spatial resolution of ~0.24 nm defined by the scanning step size of the electron probe. The position specific diffraction was then averaged over regions of 3x3 and 5x5 to illustrate the change in the projected displacement vector measured at larger scales. In FIG.5A, top-left pointing displacements colored by blue and top right pointing displacement colored by red are seen dominating the polarization and discretely populated over the mapped area, with a few domains in the size of 1-2 nm showing collective polarization toward the same direction. Over all, both top-left and top-right polarization featured by offset 020 components and mono-directional 002 components indicate the dominance of rhombohedral phase in the projected length scale of a unit cell. The abrupt transitions in the displacement vector $\theta$ occur between adjacent probe positions indicates abrupt transitions between rhombohedral cells. These transitions also occur on the length scale of individual unit cells.  To approximate the effects of using larger probe size to map the polarization, the data was binned in 3x3  (\ref{fig:empad}.B) and 5x5 (\ref{fig:empad}.C) regions, effectively equivalent to a 0.8 nm and 1.3 nm FWHM probe size. At higher length scales, the abrupt changes in polarization of adjacent unit cells lose visibility, and emerge into gradual and smooth transition boundaries between red-colored and blue-colored domains with uniform polarization, eventually with the 5x5 binning, the polarization along 020 was attenuated due to the effects of averaging 020 and 0-20 rhombohedral orientations. This data is in good agreement with the observations in prior studies using large probe sizes, showing 4-10 nm continuous coordinated rhombohedral regions. Additionally the pair composition function was calculated for the data. Separating the data into majority left and majority right projected columns. This data presented in \ref{fig:mapping} D, shows a limited change in the PDF, trending towards simple cubic in the larger binned regions. 

\begin{figure}[htp]
\centering
\includegraphics[width=3.2in]{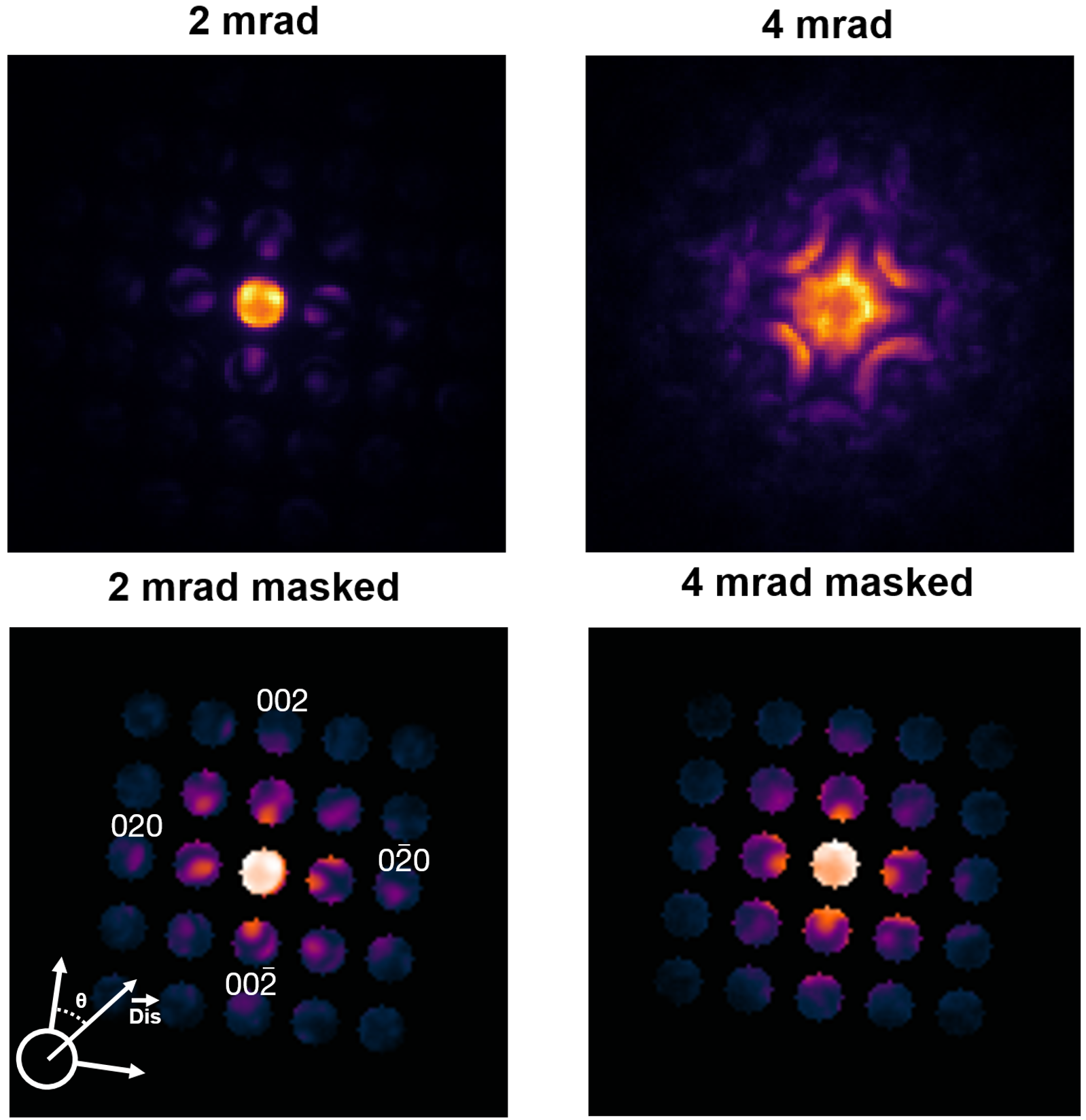}
\caption{Experimental data taken at 2 and 4 mrad convergence angles from nearby positions of a 35 nm thick sample. Samples have been masked and normalized to show the similarities in structure post masking.}
\label{fig:empad}
\end{figure}

To understand our observation of abrupt transition of rhombohedral unit cells polarized oppositely along 020, simulations of diffraction from BaTiO3 with different combination of top-left and top-right polarized rhombohedral unit cells along the thickness direction were performed to determine the extent of change in composition required to effect the change in measured displacement direction. For a range of sample thicknesses from 3 to 35 nm, the displacement vector was calculated for compositions ranging from 100 percent of one orientation to a 50 percent mix of both 002 and 00-2 composition. The relationship between the displacement vectors, thickness, and the composition are shown in Figure \ref{fig:angle_composition}.

\begin{figure}[htp]
\centering
\includegraphics[width=3.2in]{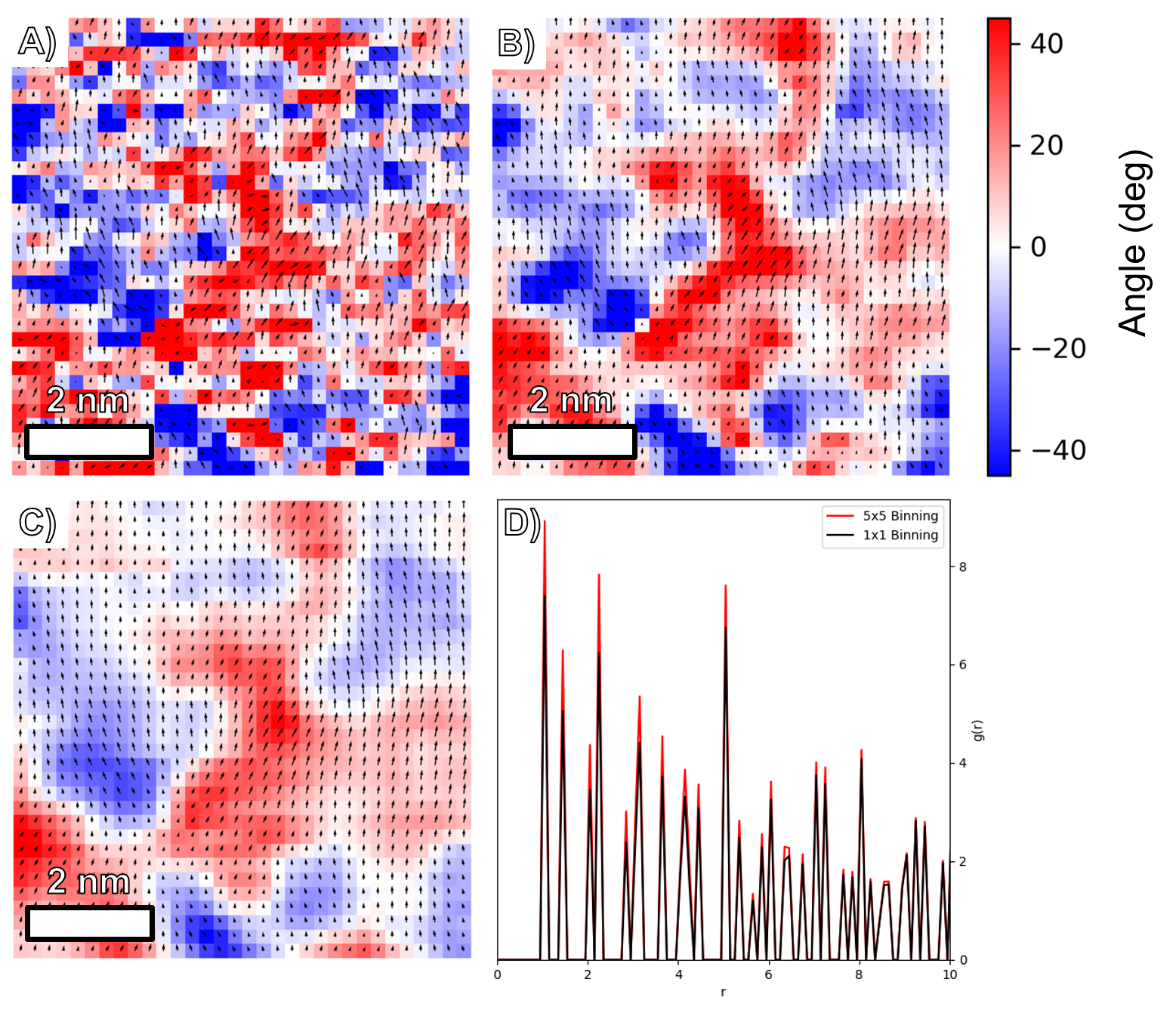}
\caption{Friedel Pair analysis results performed by averaging the patterns over A) as imaged B) 3x3 binning, and C) 5x5 binning. D) Pair correlation function results for regions with a nominally left composition.}
\label{fig:mapping}
\end{figure}

Despite the expected change in polarization measured in the simulated diffraction patterns, thickness effects lead to undependable displacement vectors at thicknesses above 26 nm and below 6 nm. Dynamical diffraction effects from samples thickness above 26 nm, and minimal diffraction contrast from sample thickness below 3 nm lead to inversion of the LR vector and unreliable measurement respectively. Comparison of PACBED simulation with the experimental PACBED allowed the determination of the sample thickness to 19 nm. Using the estimated thickness of 19 nm, further simulations to investigate the nature of diffraction generated by the beam passing through multiple displacement directions were performed.  It was observed that a beam passing through two separate regions of rhombohedral displacement vector with opposing 020 components would result in an apparent 002 polarization with tetragonal symmetry. From the 19 nm data, a change of 20 degrees in displacement vector measured in the diffraction can be caused by a change of polarization direction in as little as 10 percent of the unit cells between adjacent columns. At higher angles, a larger change is required to effect the same difference in vector angle, for example, a change of 30 percent is needed for a 20 degree shift from 45-25 degrees. Therefore, the observed dispersion of scattered rhombohedral phase with abrupt change in polarization direction across adjacent unit cells is the result of variation in distribution of the polarization displacement along the depth direction, which can be equivalent to the change of one out of ten unit cells. 

Considering both the experimental results and the simulation, beyond available studies showing the critical length scales of nms where the assembly of rhombohedral domains emerges into a tetragonal phase, this new result suggests that at the length scale of a single unit cell, the rhombohedral phase is represented by randomly polarized cells in the three dimensional volume. The change in population of the polarized rhombohedral cells is the origin of the nanosized domains and the transition between the domains.

\begin{figure}[htp]
\centering
\includegraphics[width=3.2in]{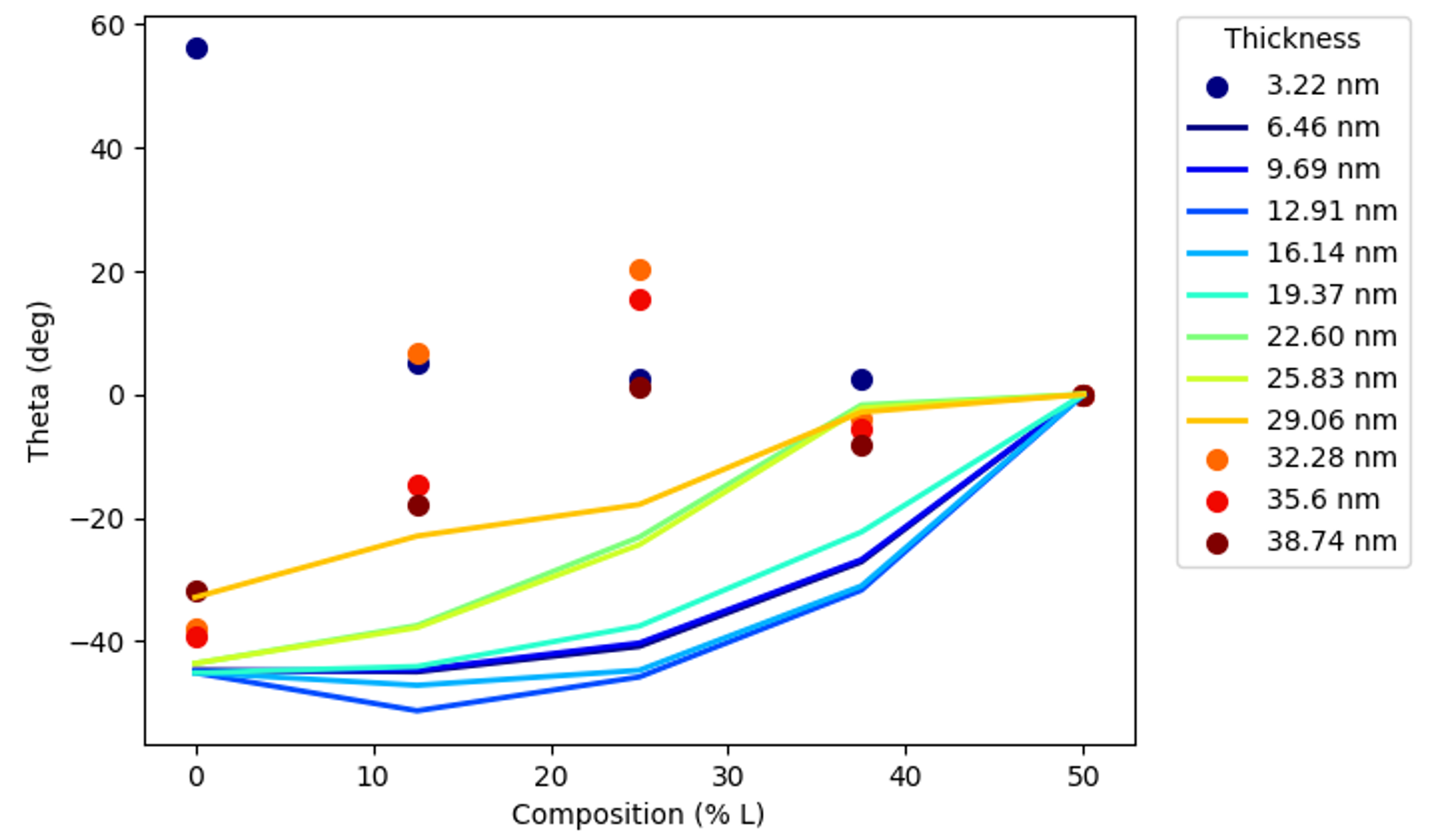}
\caption{Influence of the composition on the projected angle($\theta$) from Bloch wave simulated data at thinner thickness ranges.}
\label{fig:angle_composition}
\end{figure}

To visualize the effect of the polarization distribution among unit cells on the electric field, additional measurement on the COM of the center disc was done using a separate experimental data set acquired from the same sample. When comparing with the displacement map using Friedel pairs , the COM data shows little to no 020 component to the polarization, demonstrating less sensitivity to the deviations in the local geometry and more dependence on the overall domain polarization. This is to be expected since the electron distribution in the center beam is dependent on local electric field contributed from both local and long range Coulomb interaction, whereas the intensity distribution among diffracted discs is dependent on the local structure. Coupling the COM analysis with Friedel pair analysis using the partially overlapping CBED potentially allows decoupling the overall electric field and the structure at the unit cell level. 

\begin{figure}[htp]
\centering
\includegraphics[width=3.2in]{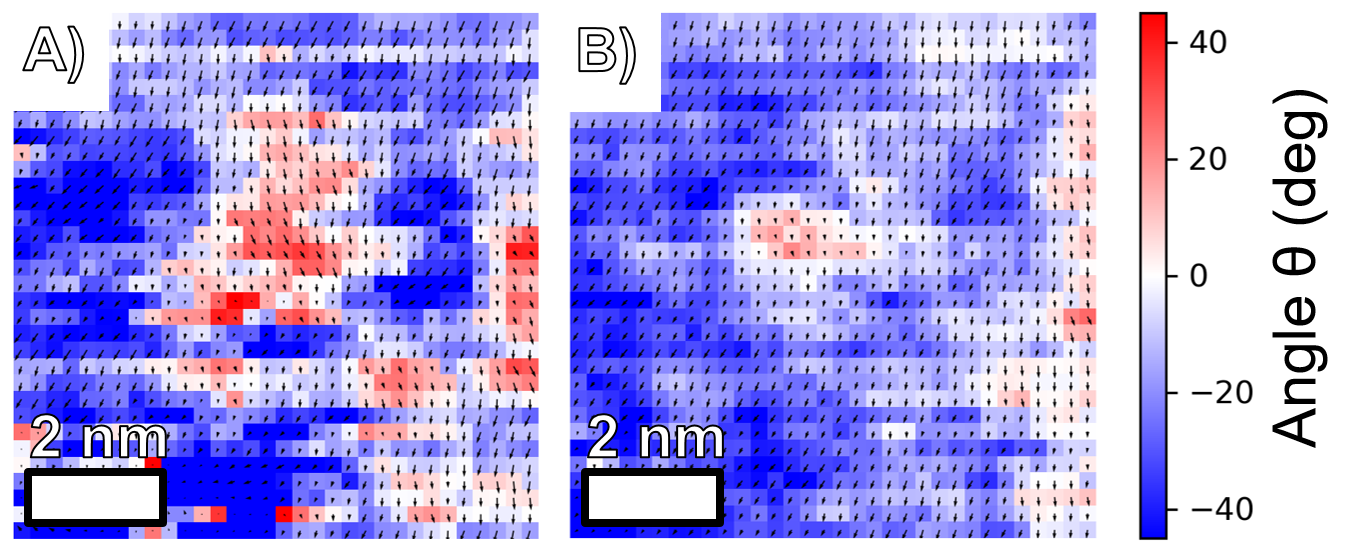}
\caption{Polarization and Orientation mapping performed on a 35 nm thick region using A) Friedel pair analysis and B) COM analysis of the masked center beam. }
\label{fig:comparitivemapping}
\end{figure}

\section{Conclusions}

We have demonstrated a method of interrogating the projected symmetry and phase of a local sample area specific to the probe position using 4D STEM data obtained on the order of a single unit cell in spatial resolution. By increasing the convergence angle and masking the overlapping regions in the diffraction, this method allows the analysis of the projected symmetry and phase at each position. As predicted by the order-disorder phase transition mechanism, we observe local correlated rhombohedral regions in a largely tetragonal domain. Within the nanoregions range from 1-5 nm in width, the results show that the polarization domains break down further to randomly polarized unit cell regions. This indicates that as the spatial resolution of the method increases, the information we can extract about the local correlation present in the rhombohedral structures decreases. This breaks down the structural observations into three critical length scales. When observations are performed at the resolution of ~1 UC, roughly correlated rhombohedral regions can be observed with local fluctuations on the order of unit cell by unit cell. When the probe sizes increase to the 0.8 to 1.3 nm regime, local fluctuations decrease, and the overall nanoregions become clearer. As the probe sizes/binned regions increase, the data trends towards a homogeneous tetragonal polarization, as predicted by previous studies. This indicates that the fluctuation of population among rhombohedral unit cells with opposite displacement along the three-dimensional volume covered by a unit cell is critical to the understanding the origin of phase and symmetry evolution across different length scales in BaTiO3.

\section{Acknowledgments}
The Authors acknowledge the assistance, advice and sample support provided by the Prof. Elizabeth Dickey's group. 

This material is based upon work supported by the National Science Foundation under Grant No. DGE-1633587.

This work was performed in part at the Analytical Instrumentation Facility (AIF) at North Carolina State University, which is supported by the State of North Carolina and the National Science Foundation (award number ECCS-2025064). The AIF is a member of the North Carolina Research Triangle Nanotechnology Network (RTNN), a site in the National Nanotechnology Coordinated Infrastructure (NNCI).

\section{References}
\nocite{*}
\bibliographystyle{ieeetr}
\bibliography{BTO_Paper}

\end{document}